\documentstyle[prl,aps]{revtex}
\begin{document}
\twocolumn[
\begin{centering}
{\large \bf Evolution of a Bose-condensed gas under variations of the
confining potential}\vspace{1.5ex}\\
Yu. Kagan$^{1}$, E.L. Surkov$^{1}$,
and G.V. Shlyapnikov$^{1,2}$\vspace{1ex}\\

{\em (1)} {\it Russian Research Center Kurchatov Institute,\\
Kurchatov Square, 123182 Moscow, Russia}\\
{\em (2)} {\it Van der Waals - Zeeman Institute, University of Amsterdam,\\
Valckenierstraat 65-67, 1018 XE Amsterdam, The Netherlands}\vspace{0.5ex}\\

\end{centering}
\begin{quote}

{\small We discuss the dynamic properties of a trapped Bose-condensed gas
under variations of the confining field and find analytical scaling
solutions for the evolving coherent state (condensate). We further
discuss the characteristic features and the depletion of this
coherent state.

}
\end{quote}
]

\vspace{4mm}

\narrowtext

The recent successful experiments on Bose-Einstein condensation (BEC)
in trapped ultra-cold alkali atom gases \cite{Cor95,Hul95,Ket95} open
a unique possibility to investigate dynamic properties of a
Bose-condensed phase. Of particular interest is the response of the
system to time-dependent variations of the confining field. In an
interacting Bose-condensed gas these properties are
non-trivial.
For example, if initially almost all trapped atoms are in
the condensate ($T=0$), then under adiabatically slow change
of the trapping potential they remain in the condensate
which now corresponds to the ground state of the system in
the instantaneous trapping field.
At the other extreme, a fast change of the potential from the
initial to final shape brings the system to an excited
superpositional (ES) state, where the admixture of the (final)
ground state can be small.
Then, even assuming complete isolation of the system from the
environment, there is the question of how the correlation
properties change.
Especially interesting are those responsible for the
reduction of the probability of inelastic processes due to the
presence of a Bose condensate \cite{Kagan85}.

Another question concerns trapped gases with negative scattering
length. The trapping field stabilizes the condensate
provided the spacing between adjacent trap levels exceeds the
interparticle interaction \cite{Ruprecht,Kagan95}.
Will this hold in the ES state or will
the system ``collapse''?

Of principal importance is the evolution of a completely isolated
many-body system which proves to be in the ES state.
Does the system undergo stochastization and imitate relaxation
behavior, at least for a
large number of particles and large interaction between them compared
to the level spacing in the potential well? This question is related
to the well-known problem of the appearence of irreversibility in a
quantum system with a large number of particles.

To answer most of the above questions we first consider the evolution of a
Bose condensate in a parabolic trapping potential $V(r)=m\omega^2r^2/2$
($m$ is the atom mass) with frequency $\omega(t)$ varing from
$\omega_0$ to $\omega_1$.
We use ideas of the analysis of the quantum motion of a particle
in a harmonic oscillator with time-dependent frequency (see \cite{Popov})
and find the solution of the time-dependent non-linear
Schr\"odinger equation for the evolving coherent state.
In certain cases our analytical
results can be compared with numerical calculations, first performed in
ref. \cite{Ruprecht}.
We further analyze the characteristic features of the evolving coherent 
state and discuss the problem of relaxation and the loss of coherence.

Let us consider a Bose gas with a fixed number of particles $N$ in
a symmetric harmonic potential with time-dependent frequency
$\omega (t)$.
We assume a pair interaction potential between atoms of the form
$U({\bf r})=\tilde U\delta ({\bf r})$. In a 3-d gas
$\tilde U=4\pi\hbar^2a/m$, where $a$ is the scattering length and
$m$ the atom mass. The Schr\"odinger equation
for the Heisenberg field operator of atoms, $\hat\Psi({\bf r},t)$,
reads
\begin{equation}             \label{Schred}
i\hbar\frac{\partial\hat\Psi}{\partial t}=-\frac{\hbar^2}{2m}\Delta\hat\Psi
+\frac{m\omega^2(t)r^2}{2}\hat\Psi +\tilde U\hat\Psi^{\dagger}\hat\Psi\hat\Psi.
\end{equation}
The field operator can be
represented as a sum of the above-condensate part and
the condensate wavefunction $\Psi_0=<\hat\Psi>$, which is a
$c$-number: $\hat\Psi=\hat\Psi^{\prime}+\Psi_0$ (see \cite{LL}).
The equation for $\Psi_0({\bf r},t)$, obtained by averaging both sides in
Eq.(\ref{Schred}), in the mean field approach has the form:
\begin{equation}  \label{Schred2}
\!i\hbar\frac{\partial\Psi_0}{\partial t}\!=\!-\frac{\hbar^2}{2m}\!
\Delta\Psi_0\!+\!
\frac{m\omega^2(t)r^2}{2}\Psi_0\!+\tilde U[|\Psi_{\!0}|^2\!+\!2n^{\prime}]
\Psi_0,\!\!
\end{equation}
where $n^{\prime}(\!{\bf r},t)\!\!=<\!\!\hat\Psi^{\prime
\dagger}\hat\Psi^{\prime}\!\!\!>$.
The mean field equation of motion for $\hat\Psi^{\prime}({\bf r},t)$
follows from Eqs. (\ref{Schred}) and (\ref{Schred2}) :
\begin{equation}             \label{Schred3}
\!\!i\hbar\frac{\partial\hat\Psi^{\prime}}{\partial t}\!=\!\!-\frac{\hbar^2}{2m}
\!\Delta\!\hat\Psi^{\prime}\!\!\!+\!\!\frac{m\omega^2\!(\!t)r^2}{2}
\hat\Psi^{\prime}\!
\!\!+\!\!2\tilde U \!\!\left[
\!(\!|\Psi_{\!0}\!|^2\!\!\!+\!\!n^{\!\prime}\!)\hat\Psi^{\prime}\!\!\!+\!\!
\frac{\Psi_0^2\hat\Psi^{\prime\dagger}}{2}\!\right]\!\!.\!\!\!\!
\end{equation}
In Eqs.~(\ref{Schred2}) and (\ref{Schred3}), due to the condition
$n\!|a|^3\!\!\ll\!1$ ($n$ is the gas density), we omitted the terms
containing anomalous averages $<\hat\Psi^{\prime}\hat\Psi^{\prime}>$.

Frequency variations change the time and distance scales in Eqs.
(\ref{Schred2}) and (\ref{Schred3}).
Let us formally introduce new operators taking this
into account (${\bf \rho}={\bf r}/b(t)$):
\begin{eqnarray}          \label{newop}
\Psi_0({\bf r},t)&=&[b(t)]^{-d/2}\chi_0({\bf \rho},\tau(t))
\exp{(i\Phi({\bf r},t))},       \nonumber\\
\hat\Psi^{\prime}({\bf r},t)&=&[b(t)]^{-d/2}\hat\chi^{\prime}
({\bf \rho},\tau(t))\exp{(i\Phi({\bf r},t))},
\end{eqnarray}
where $d$ is the dimension of the system and the phase
\begin{equation}           \label{phase}
\Phi({\bf r},t)=(mr^2/2\hbar)\{\dot b(t)/b(t)\}.
\end{equation}
The equations of motion for the operators $\chi_0$ and
$\hat\chi^{\prime}$ in new coordinate (${\bf \rho}$) and
time ($\tau$) variables take the form
\newpage
\widetext
\begin{eqnarray}
i\hbar\frac{\partial\chi_0}{\partial
\tau}\left( \frac{d\tau}{dt}b^2(t)\right) &=&-\frac{\hbar^2}{2m}
\Delta_{\rho}\chi_0+\frac{m}{2}\left[ \omega^2(t)+\frac{\ddot b(t)}
{b(t)}\right]\rho^2 b^4(t)\chi_0+\frac{\tilde U}{b^{d-2}}[|\chi_0|^2+
2\tilde n^{\prime}]\chi_0,          \label{nSchred0} \\
i\hbar\frac{\partial\hat\chi^{\prime}}{\partial
\tau}\left( \frac{d\tau}{dt}b^2(t)\right) &=&-\frac{\hbar^2}{2m}
\Delta_{\rho}\hat\chi^{\prime}+\frac{m}{2}\left[ \omega^2(t)+\frac{\ddot b(t)}
{b(t)}\right]\rho^2b^4(t)\hat\chi^{\prime}+\frac{2\tilde U}{b^{d-2}}
[|\chi_0|^2+\tilde n^{\prime}]\hat\chi^{\prime}+\frac{\tilde U}{b^{d-2}}
\chi_0^2\hat\chi^{\prime\dagger},           \label{nSchredab}
\end{eqnarray}
\narrowtext
\hspace{-5mm} where $\tilde n^{\prime}({\bf \rho},\tau)=
<\hat\chi^{\prime\dagger}({\bf \rho},\tau)\hat\chi^{\prime}({\bf \rho},\tau)>$.
Note that the choice of phase in the form (\ref{phase}) leads to the
cancellation of terms proportional to $\nabla_{\rho}\hat\chi ^{\prime}$
and $\nabla_{\rho}\chi_0$.
It is important that the phase is the same for both $\Psi_0$  and
$\hat\Psi^{\prime}$.

Let us first consider $d=2$.
We choose $\tau(t)$ and $b(t)$ such that they are
governed by the equations:
\begin{eqnarray}
\tau(t)&=&\int^tdt^{\prime}/b^2(t^{\prime}), \label{tau2} \\
\ddot b(t)&+&\omega^2(t)b(t)=\omega_0^2/b^3(t), \label{b2}
\end{eqnarray}
where $\omega_0=\omega(-\infty)$ is the initial frequency.
Then Eqs.~(\ref{nSchred0}) and (\ref{nSchredab}) are reduced
to a universal form
\begin{eqnarray}
i\hbar\frac{\partial\chi_0}{\partial\tau}\!&=\!&-\frac{\hbar^2}{2m}
\Delta_{\rho}\chi_0\!+\!\frac{m\omega_0^2\rho^2}{2}\chi_0\!+\!
\tilde U [|\chi_0|^2\!+\!2\tilde n^{\prime}]\chi_0, \!\! \label{nsSchred0} \\
i\hbar\frac{\partial\hat\chi^{\prime}}{\partial\tau}\!&=\!\!&-\frac{\hbar^2}{2m}
\!\Delta_{\rho}\hat\chi^{\prime}\!\!\!+\!\!\frac{m\omega_0^2\rho^{\!2}}{2}
\!\hat\chi^{\prime}\!\!\!+
\!2\tilde U\!\!\left[\!|\chi_{\!0}|^2\!\!\!+\!\tilde
n^{\prime}\right]\!\!\hat\chi^{\prime}\!\!\!+\!\tilde
U\!\chi_{\!0}^{\!2}\hat\chi^{\prime\dagger}\!\!\!\!.
\!\!\!\! \label{nsSchredab}
\end{eqnarray}
Eqs. (\ref{nsSchred0}) and (\ref{nsSchredab}) are universal in the sense
that in the variables $\rho,\tau$ the problem is reduced to an interacting
Bose gas in a harmonic well with constant frequency.
Once we find the solution in the initial potential
well, we know the answer at any $t$.
One should only solve the simple equation (\ref{b2}) with initial
conditions $b(-\infty)=1,\dot b(-\infty)=0$.
For example, the expression for $\Psi_0({\bf r},t)$ reads
\begin{equation}            \label{Psi02}
\!\Psi_0({\bf r},t)\!=\!\frac{1}{b(t)}\overline{\Psi}_0\!\left( \frac{{\bf
r}}{b(t)}\right)\!\exp\!\left(\!i\frac{mr^2}{2\hbar}\frac{\dot
b(t)}{b(t)}-i\mu\tau(t)\right)\!, \!\!
\end{equation}
where $\mu$ is the initial chemical potential and $\overline{\Psi}_0({\bf
r})$ is the stationary solution for the condensate wavefunction at
$t\!\!\rightarrow\!-{\infty}$.
Using the Bogolyubov transformation, generalized for an inhomogeneous
case (see, e.g., \cite{G}), on the basis of
Eqs.~(\ref{nsSchred0}), (\ref{nsSchredab})
we can describe the evolution of the spectrum and wavefunctions of
elementary excitations.

Neglecting the excitations,
Eq.(\ref{Psi02}) describes the 2-d (radial) evolution of $\Psi_0$
in long samples (axial frequency is much smaller than the
radial one) under variations of the radial frequency, as
to a first approximation one may omit the dependence of $\Psi_0$ on the
axial coordinate.

If the non-linear interaction terms
in Eqs.~(\ref{nSchred0}) and (\ref{nSchredab}) can be omitted
the above universal scaling takes place for any $d$.
This requires at least a small ratio of
interparticle interaction to the level spacing in the initial
potential: $\eta\!\!=\!\!n_0\tilde U\!/\hbar\omega_0\!\!\ll\!\!1$, where
$n_0\!\!=\!\!|\Psi_0(0,\!-\!\infty)|^2$ (see below).

In the 3-d case for arbitrary $\eta$ there is no universality.
On the other hand, in the limit $\eta\gg 1$, neglecting the
excitations, Eq.(\ref{nSchred0}) can be again reduced to
a universal form. In this case the kinetic energy term in
Eq. (\ref{nSchred0}) is

\pagebreak
\mbox{}\vfill\mbox{}
\vspace{13mm}

\mbox{}\hrulefill\mbox{}
\linebreak
comparatively small and can be omitted (see below).
Then, introducing, instead of (\ref{tau2}) and (\ref{b2}),
the equations
\begin{eqnarray}
\tau(t)&=&\int^tdt^{\prime}/b^3(t^{\prime}),  \label{tau3}  \\
\ddot b(t)&+&\omega^2(t)b(t)=\omega_0^2/b^4(t),  \label{b3}
\end{eqnarray}
Eq.(\ref{nSchred0}) is
transformed to
\begin{equation}            \label{chi03}
i\hbar\frac{\partial\chi_0}{\partial\tau}=
\frac{m\omega_0^2\rho^2}{2}\chi_0+\tilde U|\chi_0|^2\chi_0.
\end{equation}
For $t\!\!\rightarrow\!-\!\infty$ Eq.(\ref{chi03}) has a solution
$\chi_0(t)\!=\!\overline{\chi}_0(r\!)\!\exp(-i\mu t)$,
where $\overline{\chi}_0$ is given
by the well known expression \cite{Silvera,Huse}:
\begin{equation}           \label{barchi}
\overline{\chi}_0(r)=\frac{1}{\tilde U^{1/2}}\left(
\mu-\frac{m\omega_0^2}{2}r^2\right)^{1/2}\!\!\!\!\!\!\!,\,\,\,\,\,\,
r\leq\left(\frac{2\mu}{m\omega_0^2}\right)^{1/2}
\end{equation}
and zero otherwise.
As follows from Eq.(\ref{chi03}), for any $t$ the condensate
wavefunction has the form
\begin{equation}            \label{Psi03}
\!\!\Psi_{\!0}(\!{\bf r}\!,\!t)\!=\!\!\frac{1}{b^{3/2}(t)}
\overline{\chi}_0\left(\frac{r}{b(t)}\right)\!\exp\!\left(
\!i\frac{mr^2}{2\hbar}\frac{\dot b(t)}{b(t)}\!-\!i\mu\tau(t)
\!\right)\!,\!\!\!
\end{equation}
where $b(t)$ is governed by Eq.(\ref{b3}) with $b(\!-\!\infty\!)\!\!=\!\!1,
\dot b(\!-\!\infty\!)\!\!=\!\!0$.

Eqs.~(\ref{Psi02}) and (\ref{Psi03}) conserve the norm
$\!\int\!|\Psi_0({\bf r},t)|^2d^dr\!=\!N_0$,
where $N_0$ is the initial number of particles in the condensate.
Universal solutions of Eqs. (\ref{nsSchred0}) and (\ref{nsSchredab})
conserve the norm
$\int \left( |\Psi_0({\bf r},t)|^2+n^{\prime}({\bf r},t)\right) d^dr=N$.

We should emphasize that Eqs. (\ref{Psi02}) and (\ref{Psi03}) describe
a coherent evolution of $\Psi_0$. Generally speaking,
it is very different from the condensate wavefunction corresponding
to the ground state of the system in the potential well with an
instantaneous value of $\omega$ or the final value $\omega_1$,
even if $\omega(t)$ returns to the initial frequency $\omega_0$.

The time dependence of $\Psi_0$ and
normal excitations for $d=2$ (or at any $d$ for $\eta\ll 1$, neglecting the
non-linear interaction terms)
is determined by the solution of Eq.(\ref{b2}).
The latter can be found in a general case (see, e.g., \cite{Popov}).
Note that the classical equation for a harmonic oscillator with
time-dependent frequency:
\begin{equation}         \label{cl}
\ddot \xi (t)+\omega^2(t)\xi (t)=0,
\end{equation}
leads to Eq.(\ref{b2}), if one sets
$\xi(t)\propto b(t)\exp(\pm i\omega_0\tau (t))$.
With $\xi(t)\!=\!\exp(i\omega_0t)$ for $t\rightarrow\!-\!\infty$ and $t$
interpreted as a
``spatial coordinate'', Eq.(\ref{cl}) is equivalent
to the one-dimensional Schr\"odinger equation for the reflection
of a particle with ``energy'' $\omega_0^2$
from the ``potential'' $\omega_0^2-\omega^2(t)$. At times
where $\omega$ is already constant ($\omega_1$) we have
\begin{eqnarray}
\!b^2(t)\!&\equiv&\!|\xi(t)|^2\!=\!\frac{\omega_0}{\omega_1}\left[\frac{1+R}{1-R
}-
\frac{2R^{1/2}}{1-R}\cos (2\omega_1 t+\delta)\right]\!,
\!\!\label{bas} \\
\tau(t)&=&\frac{1}{\omega_0}\arctan\left[ \frac{1+\sqrt{R}}{1-\sqrt{R}}
\tan(\omega_1t+\delta/2)\right],   \label{tauas}
\end{eqnarray}
where $R$ is the reflection coefficient and $\delta$ the phase.

For slowly changing frequency
(on a time scale $\tau_0\gg\omega_{0,1}^{-1}$) the coefficient $R$
is exponentially small,
and $b=\sqrt{\omega_0/\omega_1}$. In this case the initial
condensate is adiabatically transformed to the ground state of the
system in the final trapping field, without oscillations.
If the condition $\omega_{0,1}\tau_0\gg 1$ is not valid, for at least
one of the frequencies, the scaling parameter $b(t)$ will oscillate
with a constant amplitude given by Eq.(\ref{bas}).
There is no damping of the oscillations of the condensate density,
unless relaxation is included (see below).

The instantaneous size of the evolving condensate is related to the
initial size by $r_0(t)=r_0b(t)$ ($r_0\equiv r_0(-\infty)$).
In the case of abrupt change of the frequency we have
\begin{equation}             \label{Ri}
R=(\omega_0-\omega_1)^2/(\omega_0+\omega_1)^2,\,\,\,\,\,
\delta=0,
\end{equation}
and the function $b(t)$ oscillates from $1$ to $\omega_0/\omega_1$.
For $\omega_0\gg \omega_1$ there is a large expansion of the
condensate and then compression to the initial shape.
At times $t\ll\omega_1^{-1}$ the expansion is practically free, and
Eqs. (\ref{bas}), (\ref{tauas}) yield
\begin{equation}                \label{exp}
b(t)=(1+\omega_0^2t^2)^{1/2},\,\,\,\,
\tau(t)=(1/\omega_0)\arctan(\omega_0t).
\end{equation}
If $\omega_1=0$ the compression does not occur.
We have an expanding condensate described by Eq.(\ref{exp}) at any $t$.

At times $t\gg\omega_0^{-1}$ the characteristic velocity of free
expansion $v_0=\dot r_0(t)=\omega_0r_0$.
As follows from Eq.(\ref{barchi}), for $\eta\gg 1$ the initial
size of the condensate $r_0=(2\mu/m\omega_0^2)^{1/2}$, and we have
$v_0=\sqrt{2\mu/m}$. Since in this case $\mu=n_0\tilde U$,
the velocity is determined by the interaction between
particles.
For $\eta\ll 1$ the initial size $r_0\approx l_0$, where
$l_0=(\hbar/m\omega_0)^{1/2}$ is the amplitude of zero-point
oscillations in the initial potential, and
$v_0\approx\sqrt{\hbar\omega_0/m}$ is much smaller than 
for $\eta\gg 1$.

The same picture holds for the 3-d evolution of a condensate
in the opposite limit, where initially $\eta\gg 1$, although
we should use the scaling transformation following from
Eqs.~(\ref{tau3}), (\ref{b3}).
The latter has no analog in the quantum theory of scattering and should be
solved directly.
We again obtain a periodic function $b(t)$, and oscillations
of the condensate density will be determined by Eq.(\ref{Psi03}) which was
derived assuming a large ratio of interparticle
interaction to the kinetic energy term (IK ratio) in Eq.(\ref{nSchred0}).
For $\eta\gg 1$ the IK ratio varies as $\eta^2/b(t)$
and definitely remains large if the frequency increases 
($b(t)\leq 1$).
As follows from Eq.(\ref{b3}), fast decrease of the frequency to
$\omega_1\ll \omega_0$ leads to large oscillations with
$b_{\rm max}\approx \sqrt{2/3}\omega_0/\omega_1$.
The solution (\ref{Psi03}) will be valid if $\omega_1\agt
\omega_0/\eta^2$.
At times $t\ll\omega_1^{-1}$, where the expansion is free,
Eq.(\ref{b3}) yields $b(t)\approx\sqrt{2/3}\omega_0t$
($t\gg\omega_0^{-1}$), and the
velocity of expansion $v_0\approx\sqrt{4\mu/3m}$.
For $\omega_1\ll\omega_0/\eta^2$ the IK ratio can
become small and the scaling
(\ref{b2}), instead of (\ref{b3}), should be used.
Then the profile of the condensate density will change,
but the parameter $b(t)$ determining the characteristic size of the
condensate will be very close to that following
from Eq.(\ref{b3}).

Let us now describe the evolution of a 3-d Bose-condensed gas with
negative scattering length.
The initial condensate will be stabilized by
the trapping field if $\eta\!\ll\!1$, i.e., the IK ratio is small
\cite{Ruprecht,Kagan95}.
The prime stabilization factors are the presence of the gap $\!\sim
\!\hbar\omega_0$ for one-particle excitations and the existence of a large
energy barrier for quantum fluctuations leading to collapse
\cite{Kagan95}.
Both are related to small values of $\eta$.
In the case of radial evolution of long samples the IK ratio in
Eq.(\ref{nsSchred0}) remains constant. Hence, the evolving
condensate with $a<0$ and initially small $\eta$ will
be equally stable with respect to collapse as the initial condensate.
For the 3-d evolution with $\eta\ll 1$ the IK ratio in
Eq.(\ref{nSchred0}) varies as $\eta/b(t)$ and decrease of
the frequency (expansion) makes the evolving condensate even more stable.
On the other hand, with large fast increase of the frequency (compression)
to $\omega_1\agt \omega_0/\eta$, or adiabatic
increase to $\omega_1\agt\omega_0/\eta^2$, the parameter $\eta(t)/b(t)$
strongly increases and becomes of order unity.
This can lead to instability of the condensate with respect to collapse.
The principal difference of the uniform 3-d compression from the radial
compression of long samples is attractive for comparative experiments.

The evolving coherent state described by the wavefunction
$\Psi_0({\bf r},t)$ is an ES state.
For sufficiently large and fast change of the frequency
the admixture of the final ground state in $\Psi_0$ is very small,
which raises two questions:
What happens with correlations characteristic for the static
condensate in the absence of irreversible processes, and how fast is
the depletion of the evolving coherent state.
Analyzing the first question we consider correlations responsible
for the reduction of the probability of inelastic processes
due to the presence of the condensate.
The event rate
of an $m$-body inelastic process in a homogeneous gas
$\nu_m=\alpha_mZ_m\Omega$, where $\alpha_m$ is the rate constant,
$\Omega$ the system volume, and $Z_m=<[\hat\Psi^{\dagger}(0,0)]^m
[\hat\Psi(0,0)]^m>$ the local density correlator \cite{Kagan85}.
For three-body recombination we have $m=3$, and for spin-dipole
relaxation $m=2$.
In the absence of condensate $Z_m=m!(\overline{n})^m$, where
$\overline{n}$ is the average particle density.
At $T\rightarrow 0$ in a stationary condensate the density fluctuations are
suppressed and $Z_m=(\overline{n})^m$.
Hence, $\nu_m$ in the condensate decreases by a factor $m!$ \cite{Kagan85}.

In the spatially inhomogeneous evolving Bose-condensed gas, generalizing
the above expression for the event rate we have
$\nu_m(t)=\alpha_m\int d^3{\bf r}Z_m({\bf r},t)$.
The structure of the field operators in the correlator $Z_m({\bf
r},t)$ is determined by the scaling transformation (\ref{newop}).
If almost all atoms are in the coherent state $\Psi_0({\bf r},t)$,
then both Eq.(\ref{Psi02}) and Eq.(\ref{Psi03})
lead to $\nu_m(t)=\nu_m(\!-\!\infty)(\Omega(\!-\!\infty)/\Omega(t))^{m-1}$, 
where $\nu(\!-\!\infty)$ is
the inelastic rate in the initial static condensate and the appearence
of the quantity $(\Omega(\!-\!\infty)/\Omega(t))^{m-1}$ is a trivial consequence 
of
changing the system volume: $\Omega(t)=\Omega(\!-\!\infty)b^d(t)$.
This result shows that coherent evolution retains the effect of partial
suppression of inelastic processes, characteristic
for the static condensate.
Since any loss of coherence will lead to an increasing inelastic rate,
there is an interesting possibility to study
the depletion of the evolving condensate through the measurement of the
rates of intrinsic or light-induced inelastic collisional processes.

In an isolated system the relaxation of the evolving coherent state will
be accompanied by the appearence of an effective temperature.
Assuming zero initial temperature and $\omega_1\!\ll\!\omega_0$,
a complete depletion of the condensate would lead to the effective
temperature $T_{\rm ef}\!\sim\!\mu$ for $\eta\!\gg\!1$ and
$T_{\rm ef}\!\sim\!\hbar\omega_0$ for $\eta\!\ll\!1$.
In the 3-d case the BEC transition temperature in the final potential,
$T_c\!\approx\!\hbar\omega_1N_0^{1/3}$ \cite{Bagnato}.
For $\eta\!\ll\!1$ the condition of complete depletion of the condensate,
$T_c/T_{\rm ef}\!\sim\!(\omega_1\!/\!\omega_0)N_0^{\!1/3}\!\alt\!\!1$,
is rather strong although there is an upper bound for the number of particles:
$N_0\!\ll\!l_0/a$.
For $\eta\!\!\gg\!\!1$, with $\mu\!=\!\eta\hbar\omega_0$ and
$\eta\!\sim \!(N_0a/l_0)^{2/5}$ \cite{Kagan95}, we obtain a
weaker condition which can easily be fulfilled:
$$T_c/T_{\rm ef}
\sim (\omega_1/\eta\omega_0)N_0^{1/3}\approx
(\omega_1/\omega_0)(l_0/a)^{2/5}N_0^{-1/15}\alt 1.$$

The question of relaxation and the loss of coherence 
in a quantum system with a large number of
particles has several non-trivial aspects, especially with regard to
dynamic evolution of a completely isolated system.
In the latter case we can discuss the imitation of stochastization and
the relaxation picture, although there are a number of reasons for the real
relaxation.
Thus far our analysis has assumed the mean field approach.
The relaxation can only appear beyond this approach: One should
at least consider the exact Hamiltonian and include in the analysis
the interaction terms proportional to $\tilde U^2$.

Fast decrease of the frequency to $\omega_1\ll\omega_0$ brings the
system to a high ES state of the final potential.
In the case of a large interaction ($\eta\gg 1$) the characteristic
energy of expansion is $\sim \mu$.
For maximum expansion ($b\approx \omega_0/\omega_1$) the evolution is
almost quasistationary and the size $r_0(t)$ is close to the size
of stationary states with energy $\sim \mu$ in the final potential.
The corresponding one-particle density of states $g(\mu)\!\approx\!
\mu^2/2(\hbar\omega_1)^3$ ($d\!=\!3$) will be very large,
ensuring that the spacing between
adjacent levels $\delta\varepsilon\!\sim\!\mu(\hbar\omega_1/\mu)^3$ is
much smaller than the interparticle interaction at maximum expansion
$n_{0{\rm min}}\tilde U\!\sim\!\mu (\omega_1/\omega_0)^3\!$.
The many-particle density of states grows exponentially, with the
exponent depending on the number of particles $N$.
Under these conditions even a small external influence can
lead to ``mixing'' of states and insert irreversibility.

The characteristic relaxation time in this case will be determined by
the collisional time $\tau_c\!\sim\![n\sigma v(\mu)]^{-1}\!$,
where $\sigma\!=\!8\pi a^2$ is the elastic cross section,
$v(\mu)$ the
particle velocity at energy $\mu$, and $n\!\sim\!n_{0{\rm min}}$.
Strictly speaking, $\tau_c$ represents the minimum relaxation time and
it would be  interesting to find possibilities to observe a
larger time of relaxation.
One may assume that even in the absence of any external influence
$\tau_c$ will be responsible for the formation of the
state which imitates relaxation.
The imitation can be promoted by the deviation of the external field
from harmonicity or spherical symmetry.

The characteristic time of dynamic evolution is determined by
$\omega_1^{-1}$, and we obtain a dimensionless parameter
characterizing the relaxation:
\begin{equation}      \label{rel}
(\tau_c\omega_1)\!\sim\!(l_0/a)(l_0/N_0a)^{3/5}(\omega_0/\omega_1)^2.
\end{equation}
This estimate shows the possibility of both fast
and slow relaxation.
For $\tau_c\omega_1\gg 1$ one can expect to observe osciullations of the
conensate.

For a small interaction ($\!\eta\!\approx\! N_0a/l_0\!\ll\!1$)
the characteristic energy of expansion $\sim\hbar\omega_0$, and we have
$(\tau_c\omega_1)\!\!\sim \!(l_0/a)(l_0/N_0a)(\omega_0/\omega_1)^2\gg 1$.
In this limit the relaxation, and hence the loss of coherence, is
always slower.

We acknowledge discussions with J.T.M. Walraven, T.W. Hijmans,
M.W. Reynolds, P. Zoller and S.A. Gardiner. This work was
supported by the Dutch Foundation FOM, by NWO (project
NWO-047-003.036), by INTAS and by the Russian Foundation for Basic Studies.%

\end{document}